- Using the Sub-Glass Transition Vibrational Dynamics to Predict Protein Stability in the Solid State: Fact or Fiction?
- Solid State Stability of Proteins and the PES

Maarten Batens[1†], Talia A. Shmool[2†], Jan Massant[3], J. Axel Zeitler[2], Guy Van den Mooter[1*]

[1]Drug Delivery and Disposition, KU Leuven,
Leuven, Belgium

[2]Department of Chemical Engineering and Biotechnology, University of Cambridge,
Cambridge, UK

[3]Biological Formulation Development, UCB Pharma,
Braine l'Alleud, Belgium

[†]These authors contributed equally to this work.
[*]Corresponding author; E-mail: guy.vandenmooter@kuleuven.be .

"An investigation of solid state protein stability and its link to the potential energy surface, thermodynamics and kinetics."

**The $\beta$-relaxation associated with the sub-glass transition temperature ($T_{g,\beta}$) is attributed to fast, localised molecular motions which can occur below the primary glass transition temperature ($T_{g,\alpha}$). Despite $T_{g,\beta}$ being observed well-below storage temperatures, the $\beta$-relaxation associated motions have been hypothesised to influence protein stability in the solid state and could thus impact the quality of, e.g. protein powders for inhalation or reconstitution and** **injection. However, to date, there is no comprehensive explanation in the literature which answers the question: How is protein stability during storage influenced by the $\beta$-relaxation? Here, we connect the shape of the potential energy surface (PES) to data obtained from our own (storage) stability study, to answer this question. The 52-week stability study was conducted on a selection of multi-component monoclonal antibody (mAb) formulations. Solid state dynamics of the formulations were probed using terahertz time-domain spectroscopy (THz-TDS) and dynamic mechanical analysis (DMA).**

## Introduction

Solid state protein formulations, e.g. protein powders for inhalation or reconstitution and injection, with different compositions but comparable stability shortly after production can display distinctly different stability profiles following extended storage. Such formulations are predominantly amorphous and effective stabilisation of proteins requires the protein and excipients to form a single phase without undergoing phase separation during preparation or storage (*1–3*). The efficacy of the used excipients has been attributed to their potential to slow down fluctuations in local conformations of the protein, as such changes in conformation are suggested to play a role in protein aggregation upon reconstitution and hence loss of therapeutic function of the protein (*3, 4*). Even though the primary glass transition or $\alpha$-relaxation temperature is traditionally used to estimate the (storage) stability of amorphous pharmaceutical solids, it is widely acknowledged that this parameter generally correlates poorly to protein stability in the solid state (*5*).

The primary glass transition is certainly an important descriptor for the characterisation of amorphous samples. It is fundamentally linked to the slow-down in large amplitude molecular motions that occur on a relatively slow time-scale. In addition, it is well known that a faster time scale secondary transition, also referred to as $\beta$-relaxation with its own glass transition temperature ($T_{g,\beta}$) exists at temperatures below $T_{g,\alpha}$. It has been hypothesised that the faster, more localised molecular motions, i.e. those that we referred to as fast ($\beta_{fast}$) as well as Johari-Goldstein relaxations ($\beta_{JG}$), are highly important with regards to protein stability in the solid state (*5*). The reason for this link is that $\beta$-relaxations within the glass are considered to act as precursors to the $\alpha$-relaxation process. For the $\beta$-relaxation motions to take place local energy barriers on the potential energy surface (PES) must be overcome through thermal activation. Additionally, some authors discern motions within a glass due to molecular side chains as non-JG relaxations (*6*).

One framework to rationalise the $\beta$-relaxation was developed using the so-called coupling model, with results showing a successful representation of the relaxation times of both the JG- and $\alpha$-relaxations (*6*). Based on this model the concept of caged dynamics was introduced to elucidate the motions that occur when the secondary glass transition temperature is reached, which is based upon the molecular structures of viscous liquids. The coupling model asserts that molecules are caged at lower temperatures by the anharmonic intermolecular potential which dissipates when sufficient thermal energy is supplied at $T_{g,\beta}$ (*7, 8*). The cage is formed by a molecule's nearest neighbouring molecules (*9*). However, because this hypothesis assumes that only intermolecular forces are responsible for the barrier in $\beta$-relaxation, it does not account for intramolecular interactions and complex coupled internal and external motions that are able to take place more freely with increasing temperatures as the free volume increases. It also fails to consider extended intermolecular interactions as opposed to simply these with the nearest neighbours. These aspects are expected to be of particular significance in proteins, macromolecules and their formulations, where folding and intramolecular interactions are non-negligible factors in the governing of their energetics. Recent work based on entropic data utilising the principles of JG-relaxation strongly suggest that a glass that has reached $T_{g,\beta}$ represents rotational or translational motions of a small number of molecules within the bulk sample (*10, 11*), and not due to cooperative motion throughout the entire bulk of the sample. This principle can be directly translated to macromolecules such as proteins, where individual parts or subunits of the molecule can exhibit substantial motion upon heating and where a distribution of motions can be active within the bulk of the sample, so long as the local potential energy barrier has been overcome.

In previous work, we have shown that terahertz time-domain spectroscopy (THz-TDS) can be used to probe the vibrational dynamics of a number of lyophilised and spray-dried formulations and to track the mobility of bovine serum albumin (BSA) and monoclonal antibodies (mAbs) with temperature (*12*). Thus, for solid state systems, in line with the energy landscape model, $T_{g,\beta}$, is associated with the $\beta$-process and defined as the point at which the molecules have a sufficient free energy to escape a local energy well and begin exploring different conformational environments with increasing temperature. Upon reaching $T_{g,\alpha}$, associated with the $\alpha$-process as a result of further increase in temperature, the mobility of the molecules can either: (1) increase gradually, as the molecules continuously explore different conformational environments, or, (2) reach a plateau, since the molecules achieve a low energy conformation and become trapped in a deep energy minimum (Fig. 1).

In the present study, we extended the work performed by Cicerone and Soles (*13*) and investigated whether or not the initial energy barrier that has to be overcome before local mobility can occur will be higher for spray-dried mAb formulations containing small amounts of glycerol. We also aim to answer the following questions: How does this behaviour change for higher glycerol contents? How do both the onset and the rate of these motions relate to solid state protein stability? In order to answer these questions a 52-week stability study was conducted using a selection of multi-component spray-dried mAb formulations covering a broad range of trehalose, glycerol and residual water contents that were intentionally selected to achieve a broad range of mAb stability.

# Results

Overall, a relative increase in the high molecular weight species (HMWS) content over time was observed for all formulations. The aggregation rate (d HMWS/d$t$) was used as the indicator for the stability of the formulation, with high values indicating poor stability. The increase in d HMWS/d$t$ was most pronounced for formulation F3, i.e. the formulation with the highest glycerol content that did not contain any trehalose.

## Terahertz Time-Domain Spectroscopy (THz-TDS)

For all of the formulations, the change in absorption with temperature was observed to take place over three distinct thermal regions that were separated by two transition temperatures, $T_{g,\beta,\text{THz}}$ and $T_{g,\alpha,\text{THz}}$ (Table 1). All characteristic parameters of the linear fit based on terahertz analysis were clearly influenced by the excipient composition of the spray-dried mAb formulations (Fig. 2).

## Dynamic Mechanical Analysis (DMA)

For all formulations a set of characteristic parameters of the $\tan\delta$ $\beta$-transition peak was determined (Table 2). The amplitude and area (Fig. 3) of the fitted mean $\tan\delta$ $\beta$-transition peak showed a clear, positive correlation with the glycerol content of the spray-dried mAb.

# Discussion

$T_{g,\beta}$ represents the thermal energy required to overcome the lowest local energy barrier on the PES. The depth of this minimum can be characterised by the quantity of energy (heat) required to be added to the system before additional states can be explored. Given that the energy requirements to overcome the minimum at $T_{g,\beta}$ were met for all formulations stored at 313 K, it is important to consider the observed differences in solid state stability (d HMWS/d$t$) over time. Taking into account d$a_2$/d$T$ being characteristic for the shape of the surface of the energy landscape between deep minima, we therefore suggest that not only the depth and presence of energy minima, but also the roughness of the PES will dictate the solid state stability of proteins in the solid state. A rougher surface, displaying a large number of relatively shallow energy minima, present across the entire surface, as well as rough surfaces within the deeper energy wells, will cause more hindrance for the molecules to explore alternative conformational states (which are less kinetically accessible) (*14*). This impact on the process kinetics can be compared to the increased energy required to push a cart down a rough cobblestone surface versus down a smooth asphalt surface. Specifically, the changes in protein motions associated with the $\beta$-process would occur on a relatively fast time-scale due to the low-lying energy barriers between different energy wells. The high energy barriers associated with the $a$-relaxation result in slower kinetics observed for this process. Therefore, we propose that not only the onset and depth of the energy minima characterised by $T_{g,\beta}$, but also the roughness of the PES, which can be characterised by d$a_2$/d$T$, govern protein stability in the solid state.

Large conformational changes or aggregate formation above the level of dimers are highly unlikely in the solid state, as illustrated by the work of Koshari et al. (*15*). However, as outlined by Cicerone and Douglas (*5*), small conformational changes exposing hydrophobic groups could increase the propensity for aggregation upon reconstitution in a hydrophilic medium. This loss of conformational integrity is likely on the level of the tertiary structure, rather than the secondary structure generally probed using Fourier transform infrared (FTIR) spectroscopy (*3*). It is reasonable to assume that these aggregation-prone conformations will be favoured energetically as water is removed during drying and in the solid state over the native conformation, which will be favoured upon reconstitution in hydrophilic media. Due to the inherently lower mobility requirements for the small conformational changes, these are possible at $T_{g,\beta}$. This entails that changing into these energetically favourable, aggregation-prone conformations in the solid state can take place during storage. Notably, it has been widely shown in the literature that protein aggregation is ultimately driven by thermodynamics, toward a low free energy state. In the PES, a low free energy state is represented as a deep minimum. From an energy landscape perspective, with heating, at $T_{g,\beta}$ parts of the protein can mobilise between energy minima from one low-energy conformation to another. It can thus be suggested that solid state protein formulations containing more proteins in aggregation-prone conformations, resulting in a larger amount HMWS upon reconstitution, have higher energy barriers to mobility compared to formulations mainly composed of proteins in the native conformation. These aggregation-prone sub-populations are consequently trapped in energy minima of the PES and will have to overcome higher energy barriers before beginning to explore different conformational environments. The presence of these higher energy barriers is in line with the higher $T_{g,\beta}$ and $T_{g,\alpha}$ values measured for formulation F3 and, to a lesser extent, F2 (Fig. 2 (a)), which contained a higher amount of HMWS following spray-drying and reconstitution.

Fig. 2 (b) clearly shows that, despite the higher energy barriers to mobility for proteins in aggregation-prone conformations, higher glycerol contents resulted in an increase of d HMWS/d$t$. This could indicate that the lower energy barrier present for the native-state proteins in F2 and F3 was likely masked by the significant increase in activation energy required for the proteins in aggregation-prone conformations present at $t_0$. Nevertheless, it is clear that a glycerol-free trehalose matrix is more efficient in slowing down degradation processes that lead to an increase in HMWS upon reconstitution. The onset of $T_{g,\beta}$ and $T_{g,\alpha}$ can be used as indicators for the barrier to mobility, influenced by the steric hindrance and free volume availability. It can be suggested that the formulations which are more aggregated upon reconstitution have higher energy barriers to mobility in the solid state. Specifically, a formulation displaying a high level of aggregation upon reconstitution could include populations of the protein in different states, some of which are trapped in low energy structures and some of which must overcome high energy barriers before beginning to explore different conformational environments. This is in agreement with the observation that F3 show the greatest aggregation and also has the highest barrier to mobility, followed by F2, then F4 and finally F1 which exhibits the lowest $T_{g,\beta}$ and $T_{g,\alpha}$ values. It can be suggested that in F1, which does not contain glycerol, trehalose can form a hydrogen-bonded network with the mAb, immobilising and stabilising the molecules during the spray-drying process. As additional excipients are added to the system, or as the flexibility of the system is increased, the mAb would experience greater steric hindrance (due to entanglement) and a reduction in free volume. Indeed, for F2 and F3 the values of $T_{g,\beta}$ and $T_{g,\alpha}$ rise, which can be attributed to the role of glycerol in serving to increase the flexibility and mobility of molecules (*5*). Thus, the reduced aggregation rate of F1 and F4 could be related to the earlier onset value of $T_{g,\beta}$ and lower energy barrier for motion to occur. The strong hydrogen bonded network between glycerol and the mAb, which was likely formed upon spray-drying, resulted in a high initial energy barrier, and thus raised $T_{g,\beta}$, which had to first be overcome before local mobility could occur for F2 and F4. As expected, for F4, $T_{g,\beta}$ and $T_{g,\alpha}$ values fall between those of F1 and F2, suggesting that this combination of glycerol and trehalose reduced the free volume and raised the barrier for mobility for this system.

The value of d$a$/d$T$ for each region observed by THz-TDS, can be used to define the extent of mobility of the system, which is largely determined by the role of the excipient and its interactions with the mAb. An increase in the value of d$a_2$/d$T$ above $T_{g,\beta}$ is observed for F2 and F3, which indicates that glycerol increases the molecular mobility and flexibility of the system (*5*). Furthermore, the THz-TDS spectra show that above $T_{g,\alpha}$, Region 3 has the lowest gradient value for F4 and F1 followed by F2, and F3 has the highest value for d$a_3$/d$T$.

This suggests that with increasing temperature, hydrogen bonds between the mAb and a stiff matrix could reduce the conformational mobility of the mAb and increase its solid state stability at elevated temperatures. This is corroborated by a higher glycerol content resulting in an increased amplitude of the $\tan \delta$ $\beta$-transition peak (Fig. 3 (a)), which serves as a measure for increased molecular mobility. Additionally, the amount of energy released during the $\beta$-transition increased with higher glycerol content, as indicated by the increase in area under the $\tan \delta$ $\beta$-transition peak (Fig. 3 (b)) (*16, 17*). These observations are consistent with previous work which shows that low mobility is related to high formulation stability, and here F1 and F4, which exhibit the lowest mobility above $T_{g,\alpha}$, as indicated by the value of the gradient, also show the highest (storage) stability.

This work raises the question: What is required for a protein to resist conformational changes in the solid state? It can be proposed that this is the case for proteins in energetically favourable, aggregation-prone conformations, or a system which is highly stable, due to strong protein and excipient interactions. These are the systems which exhibit high values for $T_{g,\beta}$ and resist conformational changes induced by heat. Therefore, if protein molecules are able to reach a minimum above $T_{g,\alpha}$, they exist in a relatively stable conformation, compared to a system in which the protein molecules do not reach a minimum above $T_{g,\alpha}$. The presented data offer a more comprehensive explanation for the stabilising properties observed for small flexible molecules. When examining a glycerol-containing solid state protein formulation, it is critical to consider the mutual interplay between thermodynamics and kinetics of the system: the transition temperatures ($T_g$) and the absorption coefficient gradients ($d\alpha/dT$). Additionally, we suggest further work should be completed, extending the presented hypothesis on the fundamental nature of the processes discussed in this work, to the nascent field of amorphous dispersions and co-amorphous materials.

# Materials and Methods

## Materials

A humanised immunoglobulin G4 (IgG4) monoclonal antibody was provided by UCB Pharma (Braine l'Alleud, Belgium). D(+)-trehalose dihydrate and polysorbate 20 (PS20) were obtained from Sigma-Aldrich (Steinheim, Germany). Glycerol (anhydrous), L-histidine (L-his) and L-histidine hydrochloride (L-hisHCl) monohydrate were purchased from Merck (Darmstadt, Germany) and used as received.

## Spray-drying

Feed solutions were spray-dried using a Büchi B-290 Mini Spray-dryer, equipped with a 0.7 mm two-fluid nozzle, standard cyclone, standard collection vessel and the B-296 Dehumidifier (Büchi Labortechnik AG, Flawil, Switzerland). Feed solutions had a protein concentration of 50 mg mL$^{-1}$. Settings were based on the methodology previously described by Batens et al. (*18*) with the inlet air temperature set at 393 K (outlet temperature was monitored and ranged between 343 K and 353 K), inlet air flow rate at 580 L min$^{-1}$, nozzle N$_2$ flow rate at 10 L min$^{-1}$ and the solution feed rate was set at 3 mL min$^{-1}$.

Powder fractions were then collected from both the collection vessel and the cyclone, both fractions were pooled together and dispensed into 2 mL Type I, clear, tubular glass injection vials (Schott AG, Mainz, Germany) with FluroTec® rubber injection stoppers (West Pharmaceutical Services, West Whiteland Township, PA, USA). The filled, open vials were then subjected to 2 hours of post-drying at 1 mbar and 298 K in an Epsilon 2-6D freeze-dryer (Martin Christ Gefriertrocknungsanlagen GmbH, Osterode am Harz, Germany), followed by the application of a 1 bar nitrogen atmosphere during 10 minutes before stoppers were pushed downwards, ensuring a nitrogen atmosphere in each closed vial and capped with aluminium crimp seals (Adelphi Healthcare Packaging, West Sussex, UK). Final compositions of the resulting powders are summarised in Table 3. Finally, mAb samples were stored either at 278, 298 or 313 K for a 52-week stability study.

## Physicochemical Characterisation of the Solid State

### Terahertz Time-Domain Spectroscopy (THz-TDS)

Samples were prepared for terahertz time-domain spectroscopy (THz-TDS) measurements as described by Shmool et al. (*19*) and were acquired using the methodology introduced previously (*20*). The changes in dynamics of the samples were analysed by investigating the change in the absorption coefficient at a frequency of 1 THz as a function of temperature.

### Dynamic Mechanical Analysis (DMA)

The TA Q800 dynamic mechanical analysis (DMA) powder clamp, i.e. the lower tray and upper cover plate accessory available for powder measurement, was used in conjunction with a TA Instruments Q800 dynamic mechanical analyser and the 35 mm dual cantilever clamp (TA Instruments, New Castle, DE, USA). DMA measurements were performed using argon as the air bearing gas in 'multi-frequency-strain' mode at a frequency of 1 Hz and an amplitude of 10 $\mu$m. The lower tray of the TA Q800 DMA powder clamp was filled at ambient conditions after which the cover plate was firmly pressed onto the clamp by hand. After weighing the filled clamp to allow sample mass calculation using the previously determined mass of the empty TA Q800 DMA powder clamp, the clamp was then mounted into the 35 mm dual cantilever clamp and this was subsequently tightened with a torque wrench using 14 kPa of pressure. Following closing of the furnace, temperature was equilibrated at 278 K and kept isothermally for 15.00 minutes. Subsequently, temperature was equilibrated at 138 K and kept isothermally for 15.00 minutes before ramping the temperature at a rate of 2.00 K min$^{-1}$ until the *α*-transition was observed or up to a maximum temperature of 393 K was reached to avoid clamp sticking due to the sample becoming liquid or degrading.

Each formulation was measured in duplicate and, even though storage modulus ($E'$), loss modulus ($E''$) and $\tan \delta$ responses were all taken into consideration for data analysis, only the $\tan \delta$ or damping response was reported in this work since, by definition, it represents both moduli. In the equations $\sigma$ is defined as the applied stress, $s$ as the applied strain and $\delta$ as the phase lag of a sample measured with DMA. The $\tan \delta$ response is a measure for the ratio of the system's potential to dissipate energy as heat and permanently deform relative to its potential to store energy and recover from deformation. Permanent deformation is associated with increased molecular mobility and the height (and amplitude) of the $\tan \delta$ response consequently serves as a measure for this as well. Additionally, the area under the $\tan \delta$ $\beta$-transition peak is associated with the amount of energy that released during the transition and the full width at half maximum (FWHM) is a characteristic for the time distribution across the transition (*16,17*).

Following data collection, $\tan \delta$ values were normalised for sample mass. The $\tan \delta$ $\beta$-transition peaks, defined as the first transition observed in the $\tan \delta$ signal when heating from 138 K, were then fitted using the peak analyser tool of the OriginPro 8.5.0 software package (OriginLab Corporation, Northampton, MA, USA) using the fit peaks (pro) function with baseline corrected to the minimal signal value, a smoothing window size of 10, positive peaks identified as local maxima using 100 local points without the application of peak filtering or weighting and a 500 iteration fit control with a tolerance of $10^{-15}$. When no clear second peak, i.e. $\alpha$-transition peak, could be fitted, the right limit of the $\tan \delta$ $\beta$-transition peak was manually set at the minimum before the $\alpha$-transition onset. Finally, the mean ($n = 2$) of the resulting peak amplitude, FWHM, area and $T_{g,\beta,\text{DMA}}$, defined as the temperature of the peak maximum, values were calculated for each formulation.

## Powder Reconstitution

Spray-dried powders were reconstituted at 100 mg mL$^{-1}$ by injecting a calculated volume of ultrapure water through the closed vials' septa using a syringe and 18 gauge needle. Following water addition, vials were swirled gently to evenly distribute the solvent. Aside from the initial, gentle swirl, samples were not agitated. After reconstitution, the mAb concentration of the solutions was verified using UV absorbance at 280 nm using an extinction coefficient of 1.33 mL mg$^{-1}$ cm$^{-1}$.

## 52-Week Stability Study

In order to evaluate the solid state (storage) stability of the spray-dried mAb formulations, a 52-week stability study was performed. The closed vials of the spray-dried mAb formulations that were kept sealed inside a 1 bar nitrogen atmosphere were stored at 278, 298 and 313 K. At each time point (see supporting information), a number of closed vials were collected from their dedicated storage condition and placed at 278 K before analysis. The spray-dried mAb formulations were subsequently subjected to solid state characterisation or reconstituted to 100 mg mL$^{-1}$ for aggregation-based stability assessment. Solid state characterisation at each time point consisted of residual water content determination (KF), verifying sample amorphicity (XRPD and DSC), thermal (DSC) and spectral analysis (FTIR). Aggregation-based stability assessment of the reconstituted samples consisted of determining the high molecular weight species (HMWS) and large aggregate content, using SEC and DLS, respectively, and measuring the optical density at incident wavelength of 600 nm (OD600) as a measure for sample turbidity.

Following data collection for all time points, HMWS, large aggregates and turbidity were fitted linearly as a function of time using the LINEST function included in the Microsoft® Excel® 2016 software package version 16 (Microsoft Corporation, Redmond, WA, USA) in order to obtain gradient values (d HMWS/d$t$) for each formulation and storage condition. The obtained gradient values were consequently used as quantitative parameters to measure solid state mAb instability.

**Size-Exclusion Chromatography (SEC)**

Size-exclusion chromatography (SEC) was used to quantify multimeric/high molecular weight species (HMWS) (i.e. dimers, trimers, tetramers,. . . ) in the reconstituted samples. HMWS were defined as the percentage of the total area eluting before the monomer peak and are a measure for the loss of monomeric mAb. Analysis was performed using an Infinity 1260 high performance liquid chromatography (HPLC) system (Agilent Technologies, Waldbronn, Germany) equipped with a TSK-GEL G3000SW$_{XL}$ column (5 $\mu$m, 300 mm x 7.8 mm) (Tosoh Biosciences, Germany) and UV-detector (set at 280 nm). A filtered 0.2 M sodium phosphate (pH 7.0) solution, was used as the mobile phase. Samples were diluted to 5 mg mL$^{-1}$ using mobile phase and 50 $\mu$l were injected onto the column. The process was run isocratically at ambient temperature, for 15 minutes with a flow rate of 1 mL min$^{-1}$ and two injections were performed for each sample. Peaks were integrated with Empower (version 7.30.00.00) and the ApexTrack® algorithm (Waters Corporation, Milford, MA, USA).

# Acknowledgements

The authors would like to thank UCB Pharma for providing the equipment, materials and funding to make this study possible.

T.A.S. and J.A.Z. acknowledge funding from AstraZeneca UK Limited/MedImmune Limited and the UK Engineering and Physical Sciences Research Council (EP/N022769/1). T.A.S. would like to thank the AJA-Karten Trust and the AIA-Kenneth Lindsay Trust for their financial support.

# Figures and Tables

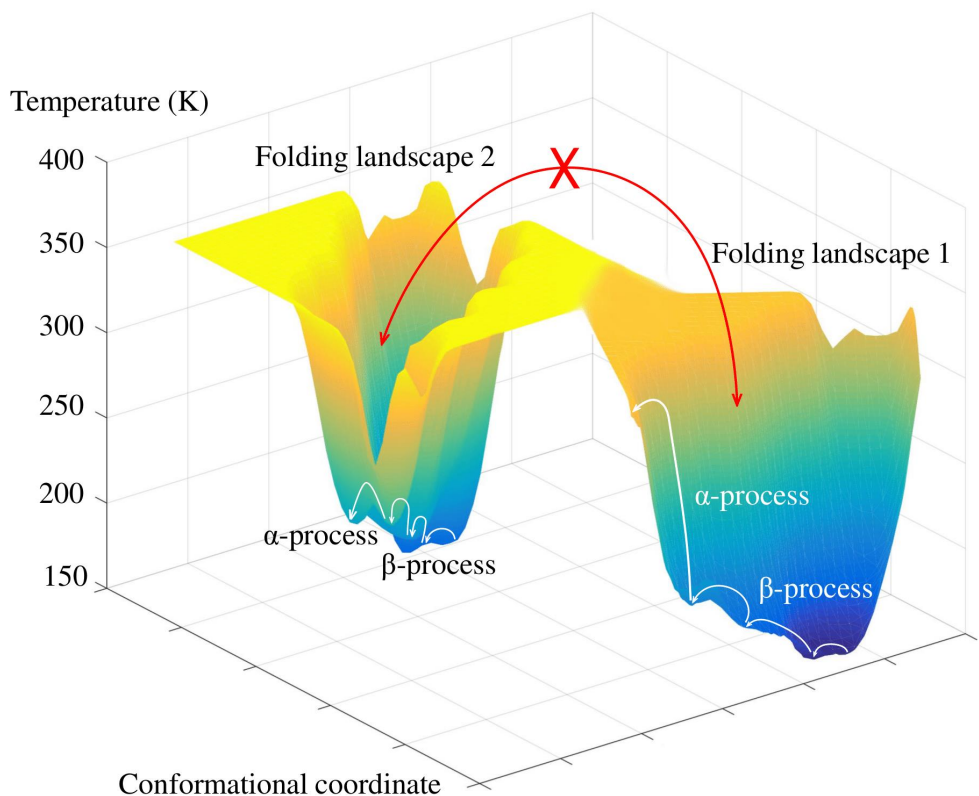

**Fig. 1. Illustration of the potential energy barrier showing the different minima between which a protein molecule can and cannot move in the solid state before denaturing.** A protein can exist in multiple conformations, and each different conformation of the protein is associated with a minimum. Each minimum is separated by energy barriers on the potential energy surface. The deepest minimum in the landscape can be linked with molecules in the most stable state. At low temperatures the protein conformation is trapped in a deep energy minimum with limited molecular mobility. With sufficient input of heat energy to the system, the *β*-process occurs and the protein can begin to move from one energy minimum to a different energy minimum and explore an ensemble of conformations, e.g. small conformational changes exposing hydrophobic groups in the solid state. With additional energy input the *α*-process occurs and the protein can overcome greater energy barriers and achieve larger conformational changes. As indicated on the illustration, the protein cannot move between different folding landscapes, e.g. between the native and a mis- or unfolded alternative state, in the solid matrix and denatures above 400 K. However, reconstitution of the protein in an aqueous environment will completely change the energetics of the landscape.

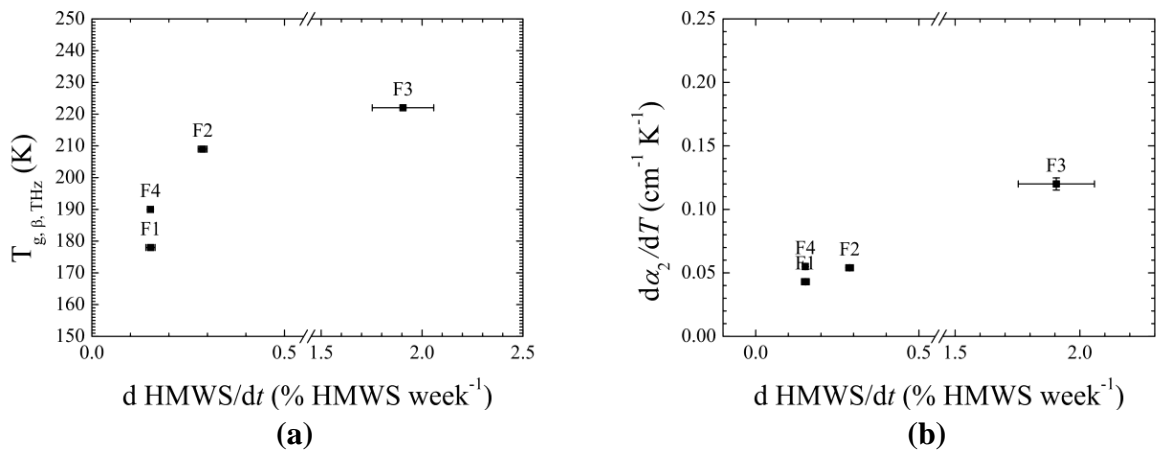

**Fig. 2. (a) $T_{g,\beta,\text{THz}}$ and (b) $d\alpha_2/dT$ as a function of d HMWS/dt for samples stored at 313 K.** Horizontal error bars depict the standard error of the slope as a measure for the precision of the regression analysis. Vertical error bars depict the standard deviation for $n = 3$ samples.

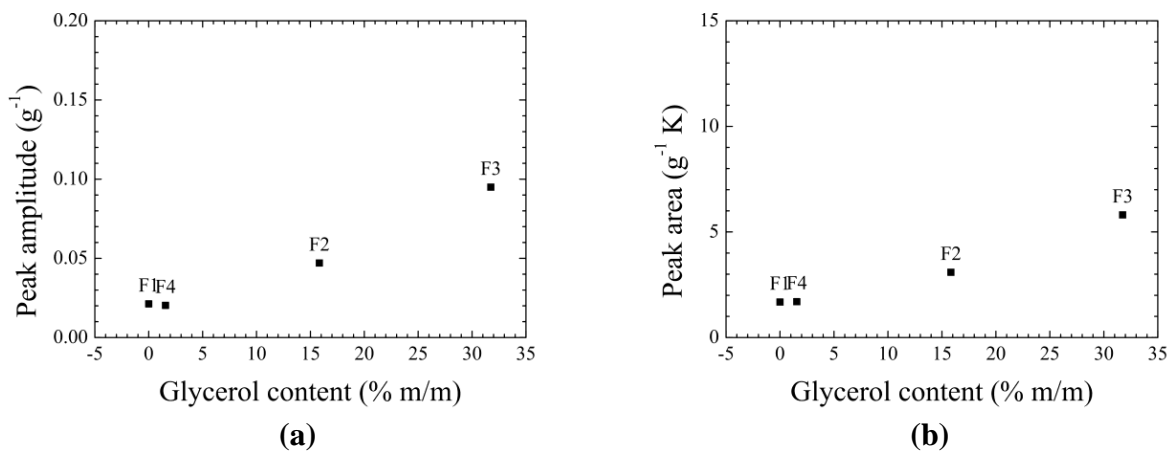

**Fig. 3. (a) The amplitude and (b) area of the fitted mean ($n = 2$) tan $\delta$ $\beta$-transition peak based on DMA as a function of the glycerol content of formulations F1, F2, F3 and F4, respectively.** Values were determined on $t_0$ samples stored at 278 K.

**Table 1. Gradients of the linear fit ($y = m\alpha + c$) for the respective temperature regions as well as the respective glass transition temperatures determined based on the terahertz analysis of spray-dried mAb formulations.** For all samples three regions were identified.

|    | d$\alpha_1$/d$T$ cm$^{-1}$ K$^{-1}$ $n=3$ | ± SD | d$\alpha_2$/d$T$ cm$^{-1}$ K$^{-1}$ $n=3$ | ± SD | d$\alpha_3$/d$T$ cm$^{-1}$ K$^{-1}$ $n=3$ | ± SD | $T_{g,\beta,\text{THz}}$ K $n=3$ | $T_{g,\alpha,\text{THz}}$ K $n=3$ |
|----|------|---------|------|--------|------|--------|-----|-----|
| F1 | 0.002 | 0.0032 | 0.043 | 0.0009 | 0.08 | 0.0035 | 178 | 268 |
| F2 | 0.016 | 0.00097 | 0.054 | 0.0010 | 0.10 | 0.0031 | 209 | 343 |
| F3 | 0.009 | 0.0032 | 0.120 | 0.0048 | 0.19 | 0.0071 | 222 | 334 |
| F4 | 0.021 | 0.0036 | 0.055 | 0.0024 | 0.04 | 0.0010 | 190 | 344 |

**Table 2. Characteristic parameters of the fitted mean ($n = 2$) tan$\delta$ $\beta$-transition peak based on DMA of spray-dried mAb formulations.**

|    | Peak amplitude g$^{-1}$ | FWHM K | Peak area g$^{-1}$ K | $T_{g,\beta,\text{DMA}}$ K |
|----|--------|-------|------|--------|
| F1 | 0.0212 | 74.47 | 1.68 | 227.17 |
| F2 | 0.0471 | 62.08 | 3.09 | 224.20 |
| F3 | 0.0950 | 57.45 | 5.81 | 232.95 |
| F4 | 0.0203 | 78.53 | 1.70 | 226.29 |

**Table 3. Composition of the spray-dried mAb formulations**

|    | mAb % m/m | Tre % m/m | L-Gly % m/m | PS20 % m/m | L-His % m/m | L-His-HCl % m/m | Water % m/m $n=3$ | ± SD |
|----|-------|-------|-------|------|------|------|------|------|
| F1 | 63.13 | 31.57 | 0     | 0.25 | 0.66 | 2.81 | 1.57 | 0.07 |
| F2 | 63.30 | 15.83 | 15.83 | 0.25 | 0.67 | 2.82 | 1.31 | 0.08 |
| F3 | 63.48 | 0     | 31.74 | 0.25 | 0.67 | 2.82 | 1.04 | 0.07 |
| F4 | 62.22 | 31.11 | 1.56  | 0.25 | 0.65 | 2.77 | 1.45 | 0.23 |

# Supplementary Materials

## Supplementary Materials for this Article Include:

Supplementary Text

Fig. S1. FTIR spectra.

Fig. S2. SEC, DLS and OD600 data as a function of time.

Fig. S3. d HMWS/d$t$ as a function of the glycerol content.

Fig. S4. $T_{g,\alpha,\text{DSC}}$ as a function of the glycerol content and d HMWS/d$t$.

Fig. S5. Mean terahertz absorption coefficient as a function of temperature at 1 THz.

Fig. S6. Characteristic THz-TDS parameters as a function of the glycerol content and d HMWS/d$t$.

Fig. S7. tan $\delta$ responses collected using DMA.

Fig. S8. Characteristic DMA parameters as a function of the glycerol content and d HMWS/d$t$.

Table S1. d HMWS/d$t$ values of samples stored at 278, 298 and 313 K.

Table S2. Mean $T_{g,\alpha,\text{DSC}}$ values.

## Supplementary Methods

### Preparation of Feed Solutions

A Vivaflow 200 Laboratory Cross Flow Cassette (Sartorius, Göttingen, Germany) was used in a tangential flow diafiltration process to exchange the mAb's storage buffer with the new formulation buffer containing 15 mM L-histidine (L-his and L-hisHCl), pH 5.5, and either 25 mg mL$^{-1}$ trehalose or 25 mg mL$^{-1}$ glycerol prepared with ultrapure water (Type 1, (Resistivity ($\rho$) $\geq$ 18.2 M$\Omega$ cm at 298.15 K). Following diafiltration, solutions were further concentrated and filtered (Polyethersulfone, 0.22 $\mu$m, Merck Millipore, Bedford MA, USA). The concentration of the mAb solutions was measured using UV absorbance at 280 nm with extinction coefficient 1.33 mL mg$^{-1}$ cm$^{-1}$ and diluted to 50 $\pm$ 0.5 mg mL$^{-1}$ with ultrapure water. Following preparation, feed solutions were stored at 278 K before spray-drying.

**Karl Fischer Titration (KF)**

The residual water content of the spray-dried mAb formulations was determined using a 831 KF Coulometer with generator electrode (without diaphragm), coupled to a 774 Oven Sample Processor (Metrohm AG, Herisau, Switzerland). Samples (20 - 30 mg) were heated to 393 K in 6 mL clear glass head space vials, which were closed using septum seals with polytetrafluoroethylene inserts (Metrohm). Water contents are expressed as mass percentages unless otherwise specified.

**X-Ray Powder Diffraction (XRPD)**

An automated X'pert PRO X-ray diffractometer (PANalytical, Almelo, The Netherlands), equipped with a Cu radiation source ($\lambda_{K_\alpha}$ 1.5418 Å, voltage 45 kV, current 40 mA), was used to determine powder crystallinity. X-ray powder diffraction (XRPD) measurements were conducted in transmission mode using Kapton$^®$ (DuPont, Wilmington, Delaware) film. Samples were spun at a rate of 4 seconds per rotation with a counting time of 400 seconds per step and a step size of 0.0167 ° in the range from 4 ° to 40 ° 2$\theta$. Data was analysed using version 1.7 of the Data Viewer software (PANalytical).

**Differential Scanning Calorimetry (DSC)**

To determine the calorimetric glass transition temperature ($T_{g,\alpha,\text{DSC}}$), a DSC 3+ differential scanning calorimeter (Mettler Toledo, Columbus OH, USA) was used to determine $T_{g,\alpha,\text{DSC}}$, which was defined as the onset temperature of the step determined with a tangential baseline, as determined by version 16.10 of the STARe Evaluation Software (Mettler Toledo). For each sample, 2 - 10 mg were placed in a 40 $\mu$l standard aluminium crucible and closed using a standard lid (Mettler Toledo) under ambient conditions. Samples were subsequently heated at a rate of 10 K min$^{-1}$ from 233 or 153 to 353 K, for F1 and F4 or F2 and F3, respectively. For $t_0$, measurements were done in triplicate, for future time points, a single measurement was performed unless no clear glass transition could be determined using the obtained thermogram.

**Fourier Transform Infrared Spectroscopy (FTIR)**

Changes in the secondary structure of the mAb in the solid state were assessed based on relative changes in the positions of peaks in the 1700 - 1600 cm$^{-1}$ and 1580 - 1520 cm$^{-1}$ range, i.e. the amide I and amide II bands, respectively. Spectra were recorded (64 scans) in the 400 to 4000 cm$^{-1}$ range with a spectral resolution of 2 cm$^{-1}$ using a Vertex 70 Fourier transform infra-red (FTIR) spectrometer (Bruker, Billerica, MA, USA). Data was pre-processed and processed with version 7.5 of the OPUS software package (Bruker). Data pre-processing was performed on the absorbance spectra and consisted of atmospheric compensation ($CO_2$ and $H_2O$), rubber band baseline correction (128 points, with exclusion of $CO_2$ bands) and vector normalisation, all of which were performed using the entire measured wavenumber range.

**Dynamic Light Scattering (DLS)**

A Wyatt Möbiu$\zeta$ Zeta Potential and DLS detector (Wyatt, Santa Barbara, CA, USA) was used to quantify the fraction of large aggregates (diameter range 2.0 nm to 2.0 $\mu$m) present in the samples after reconstitution. The laser wavelength was 532 nm, detector angle 163.5 ° and samples were diluted to 10 mg mL$^{-1}$ using a 60 mM L-histidine buffer. Data were recorded at 298 K, using auto-attenuation and the peak radius cut-off set at 2.0 - 2000.0 nm to reduce buffer signals. The number of acquisitions and acquisition times were optimised for each sample, based on acquired DLS correlation curves (intensity autocorrelation as a function of time). Processing was done using the Dynamics software, version 7.3.1.15 (Wyatt).

**Turbidimetry (OD600)**

Optical density at incident wavelength of 600 nm (OD600) was taken as a measure for the turbidity of reconstituted samples, an additional criterion for the presence of aggregates exceeding the inherent size limits of SEC and DLS. Measurements were performed using a Spectramax M5 multi-detection microplate reader combined with the SoftMax Pro software package version 7.0.3 GxP (Molecular Devices, Sunnyvale, CA, USA).

## Supplementary Results
### Stability Study

All spray-dried mAb formulations were both XRPD and DSC amorphous during the entire 52-week stability study. Further solid state characterisation at each time point did not reveal any trends over time for the calorimetric glass transition temperature ($T_{g,\alpha,\text{DSC}}$). No clear peak shifts were detected over time for the amide I or amide II bands (Fig. S1). There were no changes in reconstitution time upon storage. F3 consistently exhibited the longest reconstitution time. The large variability in the reconstitution time was likely a consequence of the nature of the method that relies on qualitative visual observation, as well as the intrinsic variability due to their particulate behaviour, e.g. particles sticking together or to the vial wall, and the increasing difficulty of correctly assessing reconstitution time end points as sample turbidity increased.

With regard to aggregation-based stability assessment, Fig. S2 summarises the HMWS

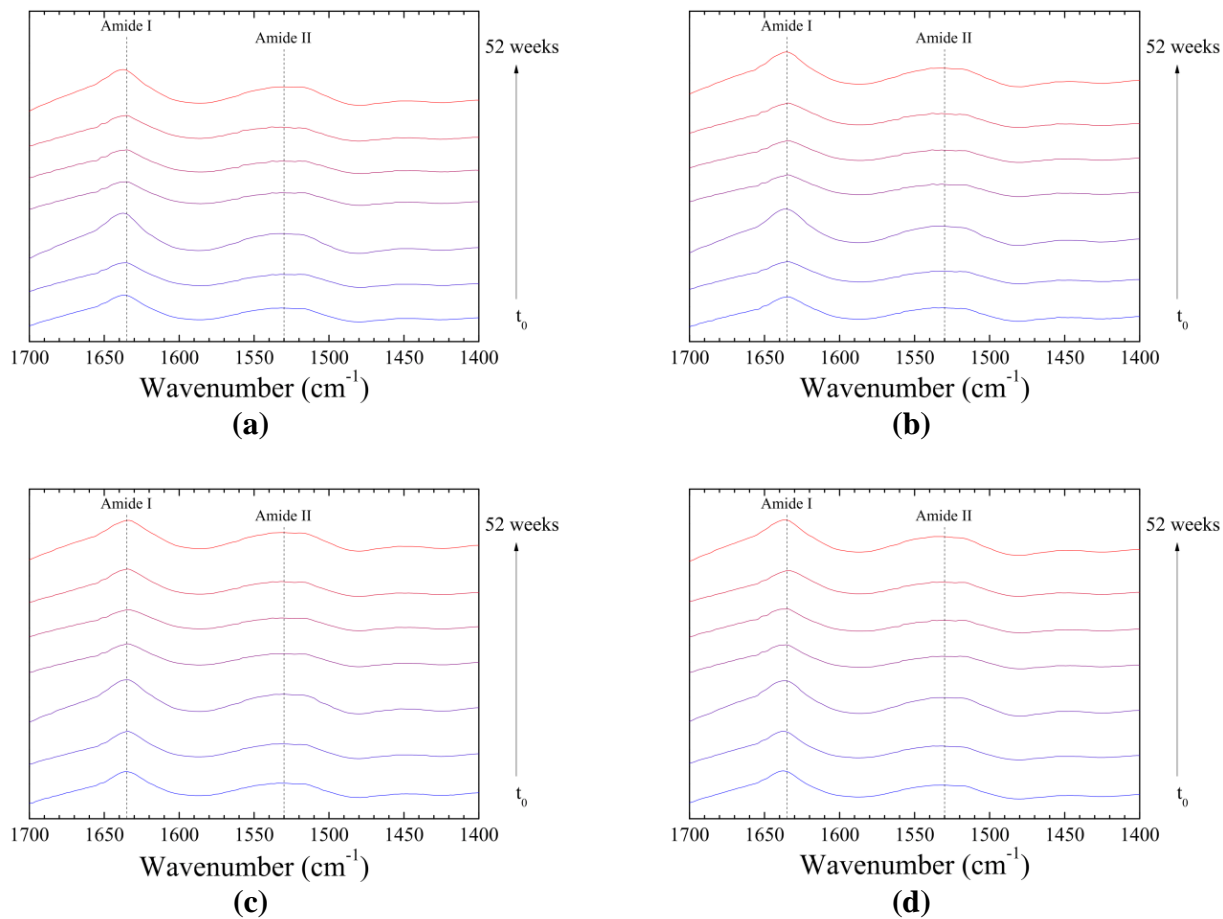

**Fig. S1. FTIR spectra collected on solid state samples of formulations (a) F1, (b) F2, (c) F3 and (d) F4 stored at 313 K during 52 weeks stacked by Y Offsets to ensure that each spectrum can be viewed clearly.** Dashed vertical lines at 1635 cm$^{-1}$ and 1530 cm$^{-1}$ serve as reference lines for the amide I (1700 cm$^{-1}$ - 1600 cm$^{-1}$) and amide II absorption band (1580 cm$^{-1}$ - 1520 cm$^{-1}$), respectively.

content, large aggregates and turbidity of the formulations stored at 313 K as a function of time. Overall, a relative increase in the HMWS over time was observed for all formulations, with the increase over time being the most pronounced for formulation F3, i.e. the formulation with the highest glycerol content and no trehalose. In this regard it should be noted that for F3, at the 39 weeks time point, the recovery, expressed as the total area under the curve as a measure for the amount of protein that passed through the column, was only about 12 % of the recovery at $t_0$. This indicated that the majority of mAb species in the sample were likely too aggregated to pass through the SEC column and withheld by the guard column. This hypothesis was further substantiated by the fact that F3, stored at 313 K for 52 weeks, was too viscous to be injected in the column following reconstitution. To avoid having to omit the 39 weeks time points for formulation F3 stored at 313 K, measurements for this time point were corrected for the low recovery by adding the area of the unrecoverable fraction relative to $t_0$ as HMWS. The validity of this correction was corroborated by a substantial increase in the coefficient of determination ($R^2$) compared to the original data set, since the uncorrected % area HMWS was lower than the previous time point, breaking the linear trend. Additional confirmation of validity was given by assessing the DLS and turbidity data for the 313 K storage condition. Data of the linear fit for the HMWS content as a function of time for samples stored at 313 K is summarised in Table S1. Large aggregates and turbidity data could not be fitted linearly as a function of time.

Increasing glycerol content resulted in a non-linear increase in $d\,HMWS/dt$ (Fig. S3). In this context it should be noted that there was little to no difference between the rate of increase in HMWS of formulations F1 and F4, containing no glycerol and 1.56 % (m/m) glycerol, respectively, despite the presence of a higher total amount of potentially plasticising species, i.e. water and glycerol (*13, 21*), in formulation F4 as formulations F1, F2, F3 and F4 all had similar residual water contents.

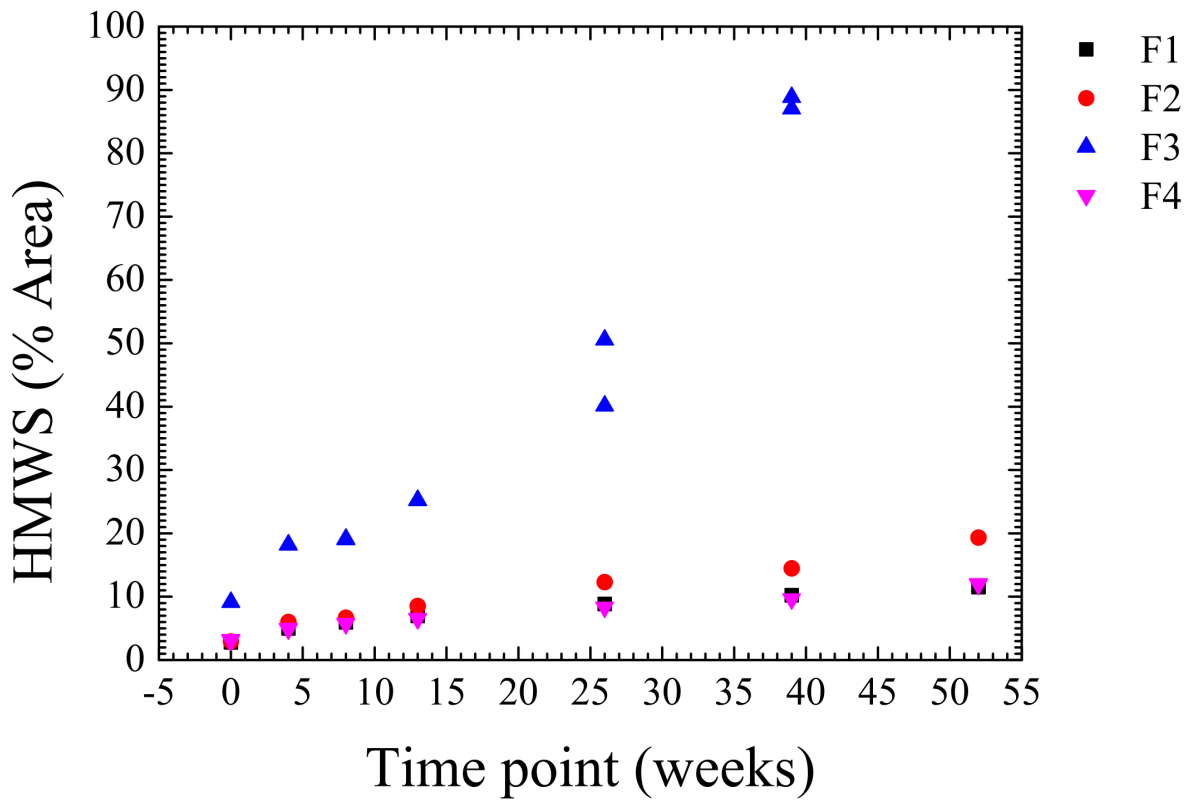

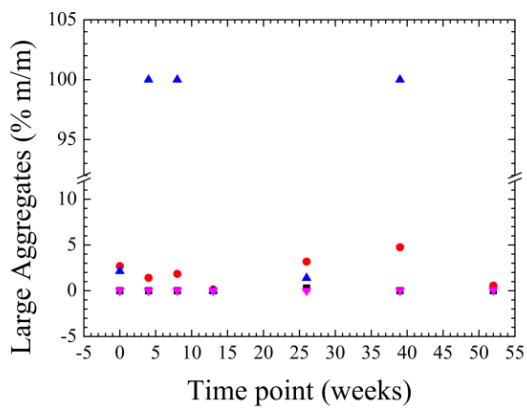

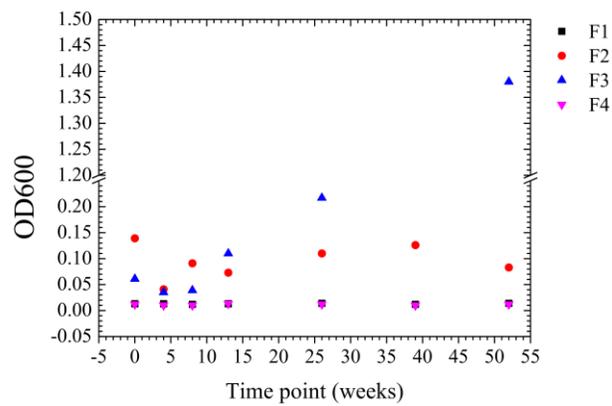

**Fig. S2. (a) HMWS content based on size exclusion chromatography analysis, (b) large aggregate content based on dynamic light scattering analysis and (c) optical density at an incident wavelength of 600 nm as a function of time following reconstitution of samples stored in solid state at 313 K during 52 weeks.**

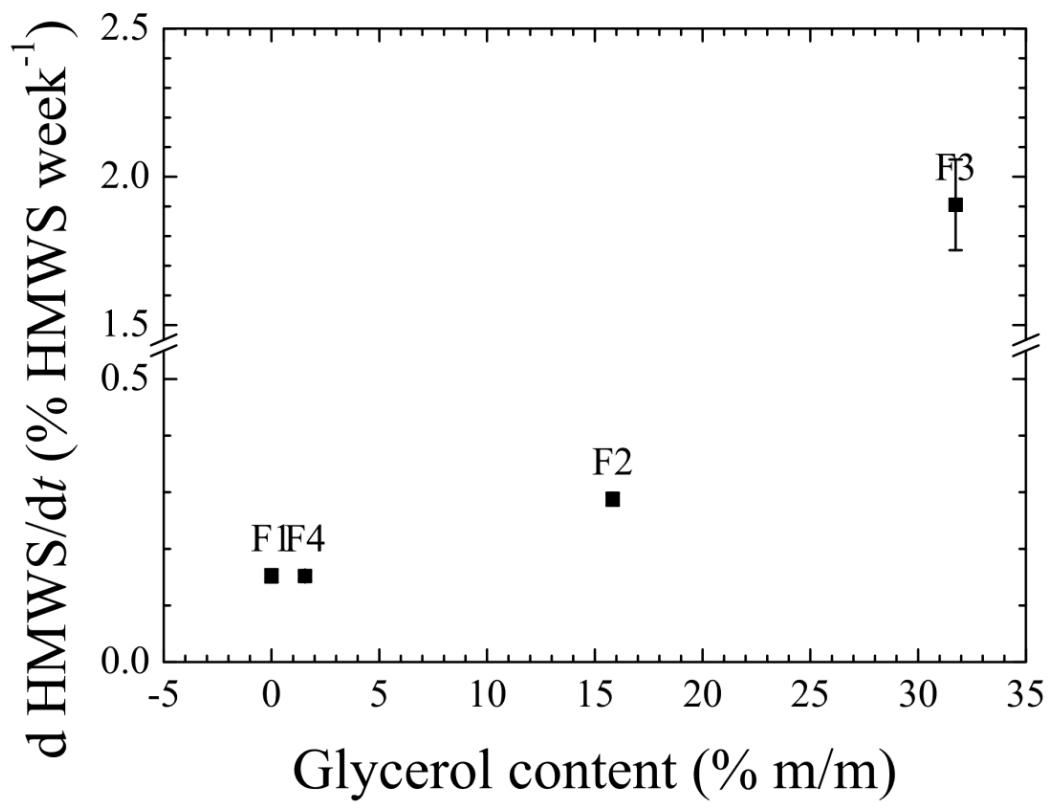

**Fig. S3. Gradient, d HMWS/d*t* for samples stored at 313 K as a function of the glycerol content of formulations F1, F2, F3 and F4.** Error bars represent the standard deviation for *n* = 3 samples.

**Table S1. Gradient, d HMWS/d$t$, and the coefficient of determination ($R^2$) of the linear fit ($y = mx + c$) for the HMWS content as a function of time.** Measurements for the 13 weeks storage of F4 at 278 K samples were identified as outliers and omitted. Measurements for 39 weeks storage of F3 at 313 K were corrected for low recovery, while F3 samples at the 52 weeks at 313 K could not be injected into the SEC column.

| Storage condition | | d HMWS/d$t$ % HMWS week$^{-1}$ | ± SE | $R^2$ |
|---|---|---|---|---|
| 278 K | F1 | 0.0115 | 0.0008 | 0.9656 |
| | F2 | 0.0095 | 0.0014 | 0.8539 |
| | F3 | 0.0483 | 0.0104 | 0.7288 |
| | F4 | 0.0082 | 0.0008 | 0.9495 |
| 298 K | F1 | 0.0497 | 0.0034 | 0.9480 |
| | F2 | 0.0502 | 0.0055 | 0.8724 |
| | F3 | 0.1178 | 0.0190 | 0.7617 |
| | F4 | 0.0409 | 0.0046 | 0.8696 |
| 313 K | F1 | 0.15 | 0.0118 | 0.9323 |
| | F2 | 0.29 | 0.0116 | 0.9810 |
| | F3 | 1.91 | 0.1527 | 0.9397 |
| | F4 | 0.15 | 0.0075 | 0.9713 |

**Relaxation Dynamics: Differential Scanning Calorimetry (DSC)**

Table S2 summarises the calorimetric glass transitions for the spray-dried mAb formulations, with samples being stored at 278 K before measurement. The mean $T_{g,\alpha,\text{DSC}}$ decreased with increasing glycerol content for the spray-dried mAb formulations. There was only a relatively small difference in mean $T_{g,\alpha,\text{DSC}}$ between formulations F1 and F4. Similarly, $T_{g,\alpha,\text{DSC}}$ decreased with increasing d HMWS/d$t$ (Fig. S4), a trend reminiscent of the one observed in Fig. S3.

**Table S2. Mean ($n = 3$) $T_{g,\alpha,\text{DSC}}$ values determined on to spray-dried mAb formulations stored at 278 K before measurement.**

| | $T_{g,\alpha,\text{DSC}}$ K | ± SD |
|---|---|---|
| F1 | 316 | 1.4 |
| F2 | 248 | 2.4 |
| F3 | 219 | 3.0 |
| F4 | 318 | 2.7 |

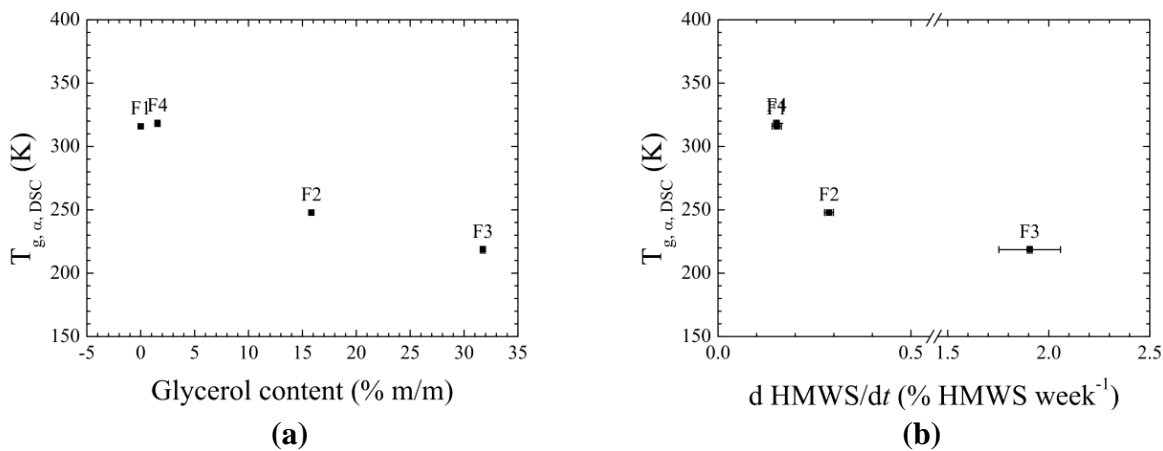

**Fig. S4. (a) Mean (n=3) $T_{g,\alpha,DSC}$ as a function of the glycerol content and (b) dHMWS/d$t$ for samples stored at 313 K of formulations F1, F2, F3 and F4. $T_{g,\alpha,DSC}$ values were determined on $t_0$ samples stored at 278 K.** Horizontal error bars depict the standard error of the slope as a measure for the precision of the regression analysis. Vertical error bars depict the standard deviation for $n = 3$ samples.

**Relaxation Dynamics: Terahertz Time-Domain Spectroscopy (THz-TDS)**

The changes in absorption at a frequency of 1 THz with temperature for spray-dried mAb formulations are plotted in Fig. S5. For all of the formulations, this change in absorption with temperature can be observed to take place over three distinct regions and two transition temperatures, $T_{g,\beta,THz}$ and $T_{g,\alpha,THz}$ (see Table 1). As the majority of motions are restricted/frozen below the $T_{g,\beta}$, d$a_1$/d$T$ was not in scope for the present study since the focus was on the onset of motion.

All characteristic parameters of the linear fit based on terahertz analysis were clearly influenced by the composition of the spray-dried mAb formulations. A clear, positive correlation is present between the glycerol content and $T_{g,\beta,THz}$ (Fig. S6 (c)). No clear correlations could be observed with $T_{g,\alpha,THz}$ (Fig. S6 (d)).

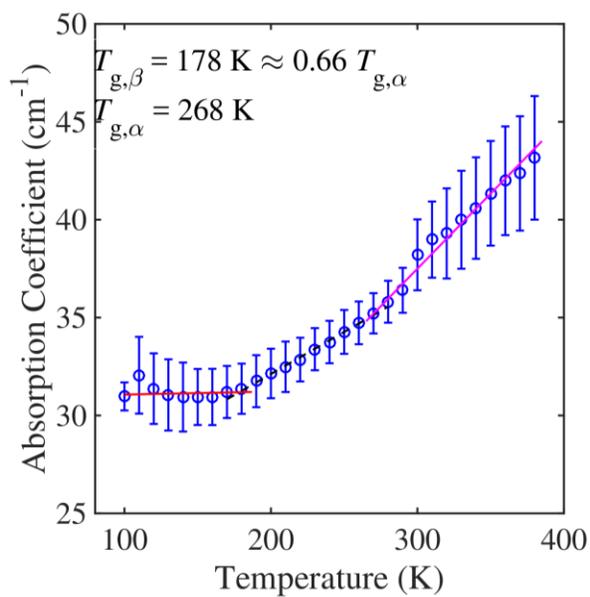
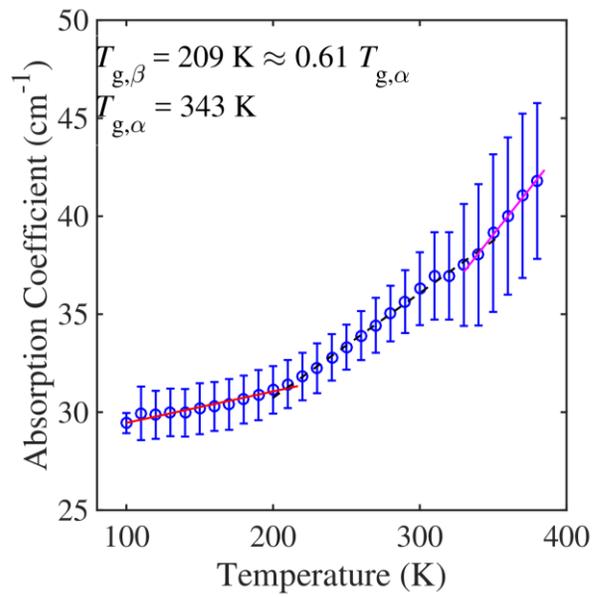
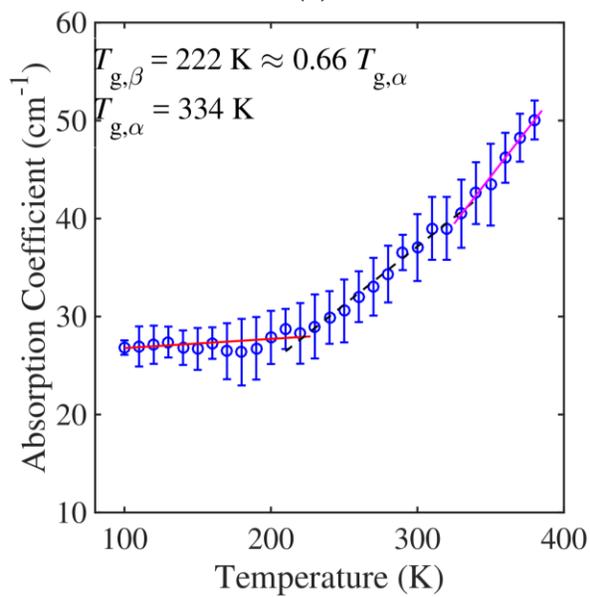
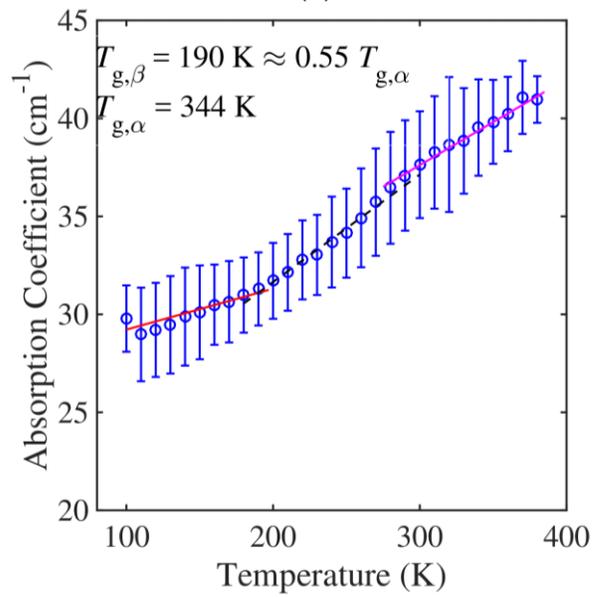

**Fig. S5. Mean terahertz absorption coefficient as a function of temperature at 1 THz for formulations (a) F1, (b) F2, (c) F3 and (d) F4. Lines show the different linear fits for the different regions.** $T_{g,\beta,THz}$ **bars represent the standard deviation for** $n = 3$ **samples.**

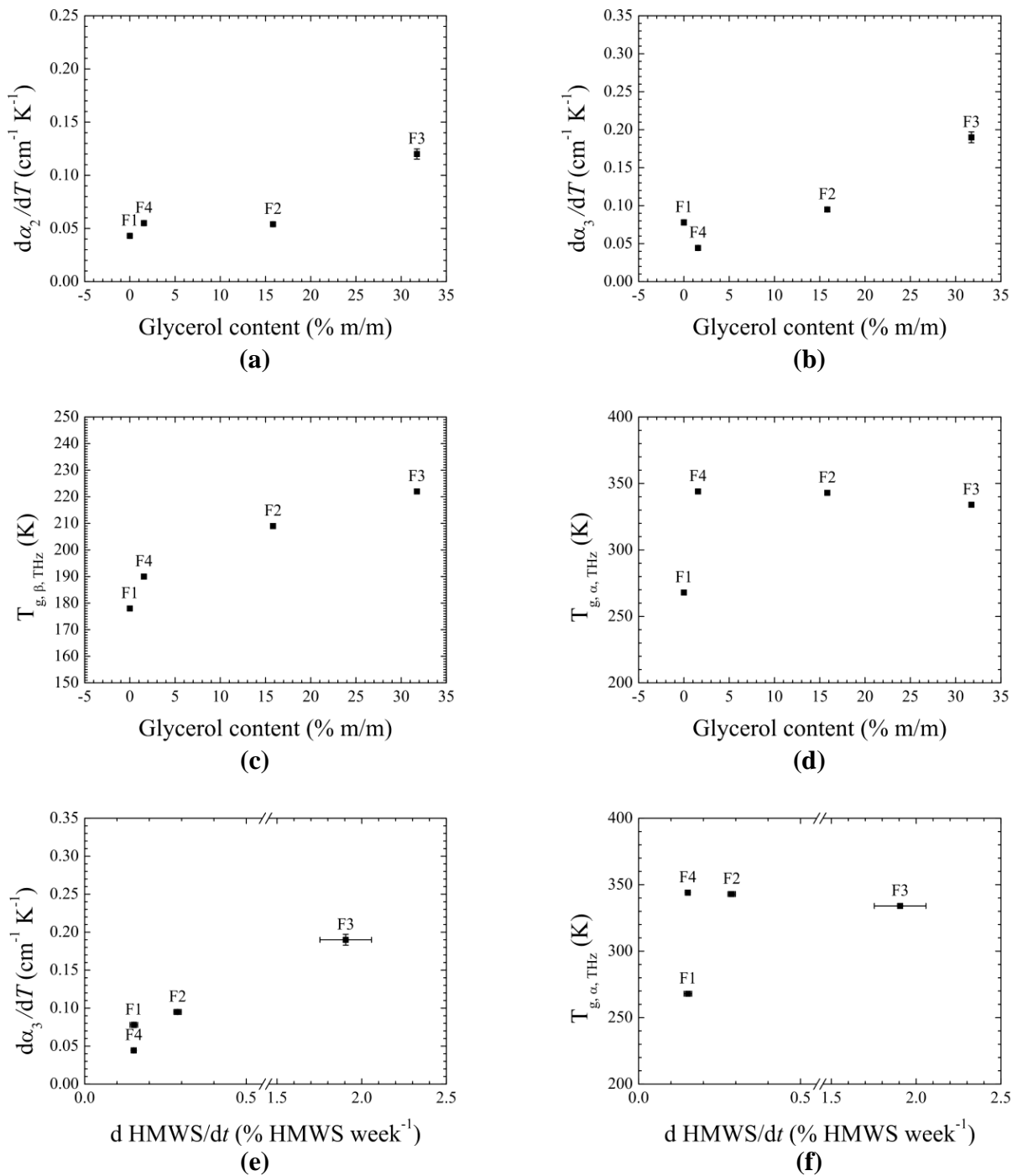

**Fig. S6.** (a) $d\alpha_2/dT$, (b) $d\alpha_3/dT$, (c) $T_{g,\beta,\text{THz}}$ and (d) $T_{g,\alpha,\text{THz}}$ as a function of the glyc- erol content of formulations F1, F2, F3 and F4, respectively, as well as (e) $d\alpha_3/dT$ and (f) $T_{g,\alpha,\text{THz}}$ as a function of d HMWS/d$t$ for samples stored at 313 K. **Vertical Error bars represent the standard deviation for $n$ = 3 samples. Horizontal error bars depict the standard error of the slope as a measure for the precision of the regression analysis.**

**Relaxation Dynamics: Dynamic Mechanical Analysis (DMA)**

Fig. S7 depicts the normalised $\tan \delta$ response for DMA measurements of each formulation. The repeat runs were found to be in good agreement with the results of the first run. Overall, two clear peaks can be observed for all formulations except F2, for which no clear $\tan \delta$ $\alpha$-transition peak could be defined. Peak deconvolution was found to generally result in a poor or no fit for the $\tan \delta$ $\alpha$-transition peaks, while good fits were obtained for the $\tan \delta$ $\beta$-transition peaks, likely due to decreased sample stiffness with increasing glycerol content in the temperature region preceding the glass transition. As $\alpha$-transition peaks could not be resolved in a satisfactory manner, only $\tan \delta$ $\beta$-transition peaks were considered for further interpretation. A summary of the characteristic parameters of the fitted mean ($n = 2$) $\tan \delta$ $\beta$-transition peaks can be found in Table 2. We would like to highlight that $T_{g,\beta,\mathrm{DMA}}$ was defined as the temperature of the peak maximum, i.e. the transition midpoint, where $T_{g,\beta,\mathrm{THz}}$ was defined as the transition onset.

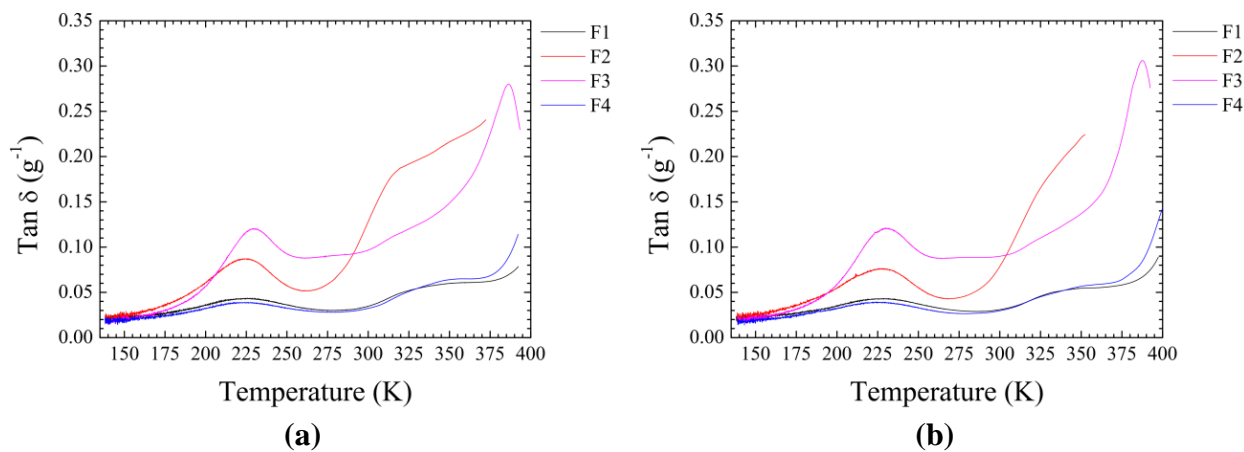

**Fig. S7.** Overlay of the tan $\delta$ responses corrected for sample mass as a function of tem- perature for all spray-dried mAb formulations based on (a) the first and (b) second DMA run of each formulation.

The amplitude and area of the fitted mean tan $\delta$ $\beta$-transition peak showed a clear, positive correlation with the glycerol content (Fig. 2) and d HMWS/d$t$ (Fig. S8) of the spray-dried mAb formulations. No clear trends were observed between the glycerol content and d HMWS/d$t$ and $T_{g,\beta,\text{DMA}}$.

Finally, as can be seen in Fig. S7, we would like to remark that the amplitude of $T_{g,\beta,\text{DMA}}$ and $T_{g,\alpha,\text{DMA}}$ peaks was generally in the same order of magnitude for a single formulation, while DMA thermograms for small molecule amorphous solids, e.g. those reported for indomethacin by Kissi et al. (*22*), show $T_{g,\alpha,\text{DMA}}$ peak amplitudes several orders of magnitude larger than those of $T_{g,\beta,\text{DMA}}$. The overall low amplitude of the $T_{g,\alpha,\text{DMA}}$ peak for spray-dried mAb formulations is in agreement with reports of hard to determine calorimetric glass transitions for pure (dry) protein formulations, as these $T_{g,\alpha,\text{DSC}}$ generally are very broad and display only a small change in heat capacity during the transition (*23*).

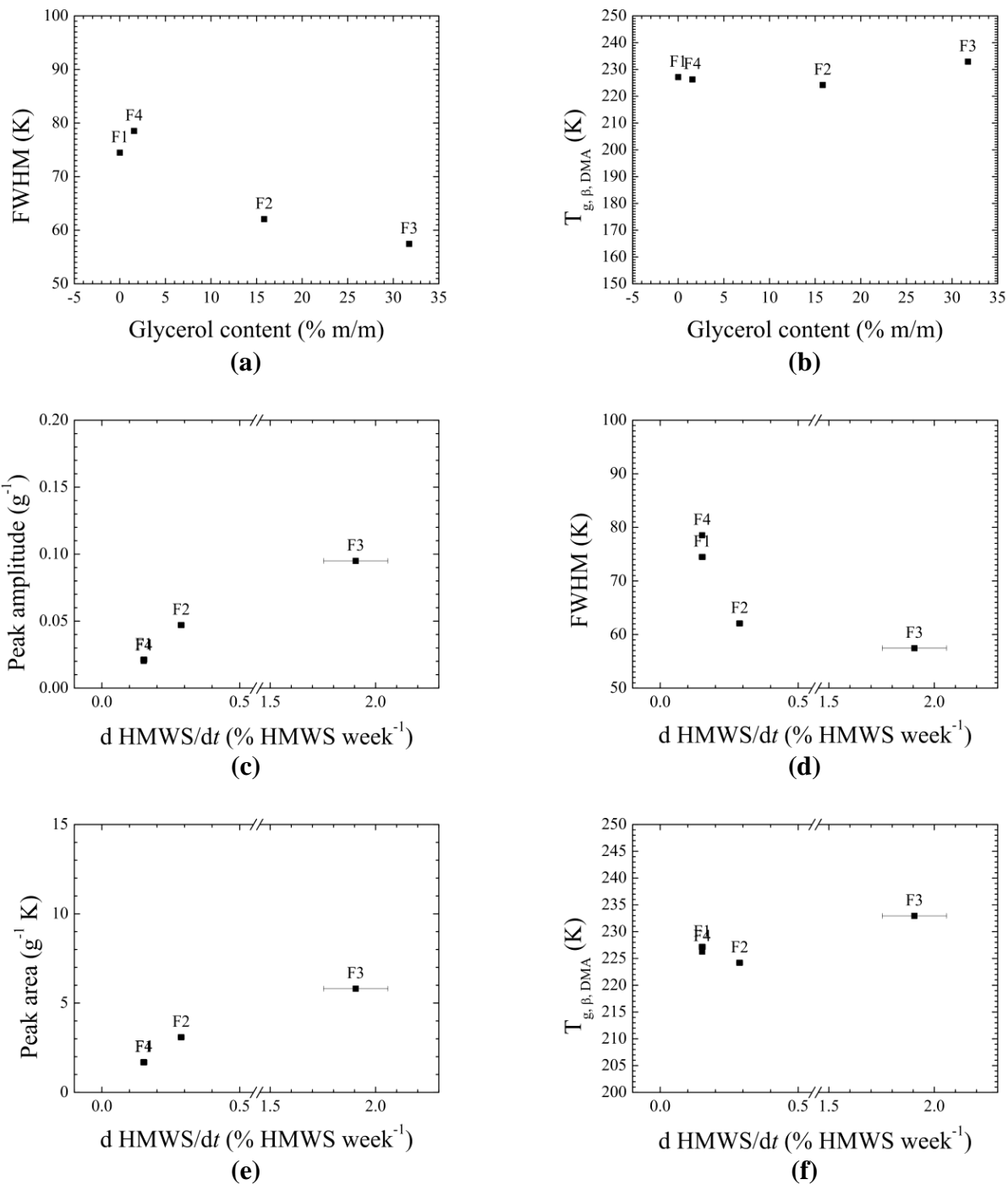

**Fig. S8. (a) The full width at half maximum (FWHM) and (b) $T_{g,\beta,\text{DMA}}$ of the fitted mean ($n = 2$) tan $\delta$ $\beta$-transition peak based on DMA as a function of the glycerol content, as well as the tan $\delta$ $\beta$-transition peak's (c) amplitude, (d) FWHM, (e) area and (f) $T_{g,\beta,\text{DMA}}$ as a function of d HMWS/d$t$ for samples stored at 313 K.** Error bars depict the standard error of the slope as a measure for the precision of the regression analysis.

36